\documentclass[12pt,a4paper]{article}
\usepackage[utf8]{inputenc}
\usepackage{authblk}
\usepackage[english]{babel}
\usepackage[T1]{fontenc}
\usepackage{amsmath}
\usepackage{amsfonts}
\usepackage{amssymb}
\usepackage{booktabs}
\usepackage{color}
\usepackage{subfigure}
\usepackage{float}
\usepackage{graphicx}
\usepackage{lmodern}
\usepackage{hyperref}
\hypersetup{colorlinks,allcolors=black}
\usepackage{fourier}

\usepackage{url}  
\usepackage[left=3cm,right=2cm,top=2.5cm,bottom=2.5cm]{geometry}
\usepackage{csquotes}
\usepackage{array}
\usepackage{cite}
\author[1,2]{Murilo S. Marques \footnote{e-mail: murilo.sodre@ufob.edu.br}}

\affil[1]{\small Instituto de Física, Universidade Federal do Rio Grande do Sul, Av. Bento Gonçalves 9500, Caixa Postal 15051, CEP 91501-970, Porto Alegre - RS, Brazil}

\affil[2]{\small Centro das Ciências Exatas e das Tecnologias, Universidade Federal do Oeste da Bahia\\Rua Bertioga, 892, Morada Nobre, CEP 47810-059, Barreiras, BA, Brazil}

\author[3]{Enrique Lomba}
\author[3]{Eva G. Noya}
\affil[3]{\small Instituto de Química Física Rocasolano, CSIC, Calle Serrano 119, E-28006 Madrid, Spain}

\author[4]{D. González-Salgado}
\affil[4]{\small Departamento de Física Aplicada, Universidad de Vigo, Campus del Agua, Edificio Manuel Martínez-Risco, E-32004 Ourense, Spain}

\author[1]{Marcia C.  Barbosa}

\title{\Large \bf Modeling the temperature of maximum density of aqueous tert-butanol  solutions}
\date{}

\begin{document}
\maketitle
	\begin{abstract}
Short-chain alcohols at high dilution are among the very few solutes
that enhance the anomalous behavior of water, in particular the value
of the temperature of maximum density. This peculiar feature, first
discovered experimentally in the early sixties, has remained elusive to
a full explanation in terms of atomistic models. In this paper, we
first introduce a two-site model of tert-butanol in which the
interactions involving hydrogen bonding are represented by a
Stillinger-Weber potential, following the ideas first
introduced  by  Molinero and Moore,  [J. Phys. Chem. B,  {\bf 113},
  4008, (2009)]. Our model parameters are fit so as to
semi-quantitatively reproduce the experimental densities and vaporization enthalpies of
previously proposed united atom and all atom OPLS models. Water is
represented using the aforementioned  potential model
introduced by Molinero and Moore, with cross interaction parameters
between water and tert-butanol 
optimized to yield a reasonable description of the experimental excess
enthalpies and volumes over the whole composition range of the
 mixture. We will see that our simple model is able
to reproduce the presence of a maximum in the change of the
temperature of maximum density for very low alcohol mole fractions,
followed by a considerable decrease until the density anomaly itself
disappears. We have correlated this behavior with changes in the local
structure of water and compared it  with the results of
all-atom simulations of water/tert-butanol mixtures.
\end{abstract}

\section{Introduction}

Aqueous binary mixtures are of great importance in science and
technology in view of the great number of (bio)chemical processes which
take place in solution. The scientific challenge posed by mixtures
stems from the extra degrees of freedom due to the composition
variables, which in turn leads to an extremely rich physical behavior
\cite{gray2011}. Among these mixtures,
water-alcohol solutions play a particular relevant role due to their
multiple uses  in various technological and everyday life  
processes, such as disinfecting and detergent agents\cite{timar1998},
solvents \cite{tereza2015,kunz2017}, dispersion
media\cite{graczova2020} and drugs constituents
\cite{vessot2012,wang2019,bakul2020}. Therefore, the study of  the thermodynamics of
aqueous alcohol solutions has attracted a good amount of research since
long \cite{franks1966,franks2013}. Having both hydrophilic and
hydrophobic functional groups, alcohols are the simplest amphiphilic
molecules. For this reason, exploring their behavior  in water solutions can provide an
invaluable insight on the role of hydrophobic and hydrophilic
interactions in biophysical processes. As a matter of fact, those are
thought to be the 
main actors in essential phenomena such as 
protein folding \cite{van2008}.  

Particular interesting are short chain monohydric alcohols that are
completely miscible with water: methanol, ethanol,
propanol and tertiary butyl alcohol (TBA).
Their solutions preserve the well known anomalous behavior of water \cite{nishikawa1989,larisa2007,subramanian2013,martina2017}
(temperature of maximum density, pressure of maximum self-diffusion
...) up to a certain concentration for which the anomalies are
destroyed, typically around  $x_{R-OH}\sim0.05$, \cite{kezic2012,bagchi2013,gomez2015}.	 Interestingly, short chain alcohols are unique among solutes in the
fact that the anomalous behavior is even enhanced \cite{wada1962a,wada1962b} by their presence in small amounts. Thus one finds that 
the temperature of maximum density exhibits an increase with respect
to that of pure water that in the case of TBA reaches a maximum value for
$x_{TBA}\sim 0.005$ \cite{wada1962a}. 
\begin{figure}[t]
  \centering
  \subfigure[\label{inter}]{\includegraphics[scale=0.35]{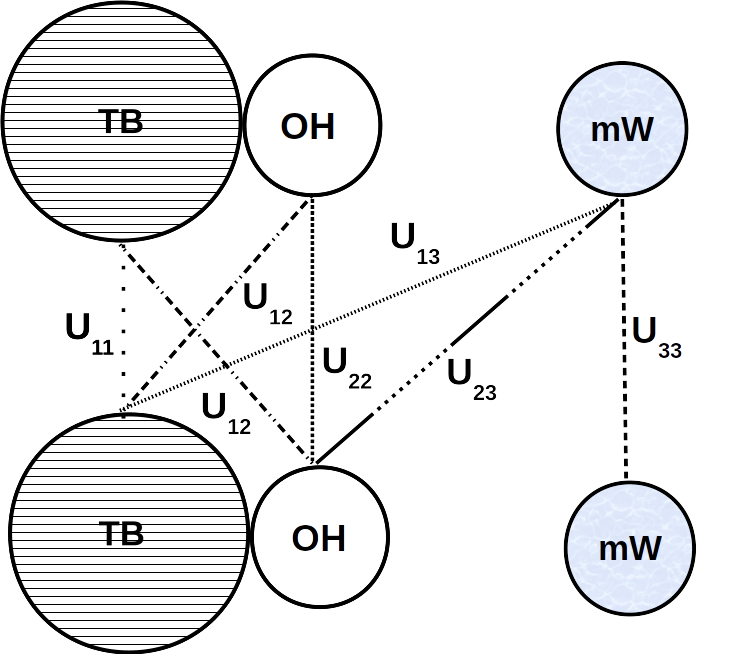}}\subfigure[\label{mol}]{\includegraphics[scale=0.5]{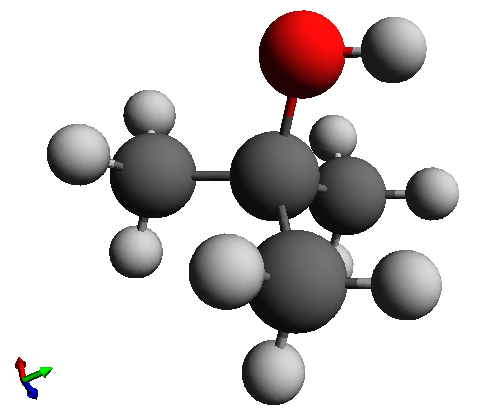}}
    \caption{(a) Our TBA and water models and their interactions. TB denotes the tert-butyl group, hydroxil group is represented by the OH site, and mW indicates a one-site water molecule modeled with the mW potential \cite{molinero2009}. (b) All-atom pictorial representation of the tert-butyl alcohol molecule.}
    \label{model}
\end{figure}

The thermodynamic anomalies of alcohol-water mixtures originate
from the corresponding anomalies of pure water
\cite{franks1972,poole1992}. Of special relevance is the volume
contraction of pure water that occurs with increasing temperature 
until a maximum density  is reached around
$4^{\circ}C$ along the atmospheric pressure isobar. The existence of
this temperature of maximum density (TMD) is probably the 
best known singularity of water,  studied already since the 17th century
\cite{bereta2000}. Its microscopic origin is based on the prevalence
of the formation of low-density ice-type structures 
over the high-density close-packed configurations right after
melting. Solutes that promote 
a more stable hydrogen-bond network would enhance the anomalous behavior
(raise the TMD), whereas those that tend to weaken it would have the
opposite effect.  This actually bring us back to the ``iceberg model''
introduced  in the forties by Frank and Evans \cite{Frank1945} to
analyze the solvation of hydrophobic solutes. According to the
``iceberg model'', the presence of a hydrophobic solute (in the case
of alcohols the alkyl chain) would induce the reorganization of the surrounding water
molecules  with an ice-like structure, which in turn would imply
an enhancement of the water anomalies (e.g. a rise in the TMD). These
solutes were originally termed 
``structure makers'', in contrast with those that tend to destroy
ice-like structures (e.g. hydrophilic groups), termed ``structure breakers''
\cite{darnel1968,hepler1969}.  This view has been supported to some
extent by the simulation study of Galamba \cite{Galamba2013}.

Since the original contributions of Wada and Umeda in the early sixties
\cite{wada1962a,wada1962b}, a number of works have addressed the issue
of the solute's influence on the TMD of water in the case of short
chain alcohols
\cite{franks1967,lilley1973,macdonald1976,kim1988,pablo2005,Su2012,lomba2016,furlan2017}. Among
these, it is worth mentioning the statistical mechanical model of Chatterjee et al.
\cite{pablo2005}.  This model predicts an increase in the TMD with the hydrophobic
character of the solute (in this particular case, molecular size) and
a decrease with the hydrophilic character (solute-solvent attraction),
in apparent agreement with the experimental results of Wada and Umeda
\cite{wada1962a}. The solute mole fraction for the maximum of the shift in the TMD is, however, larger than the experimental result. This idea proposed by  Chatterjee et al.  was later applied to a simplified dimer
 molecule where the effect of hydrogen bonding is modeled with a two
 scale potential\cite{Su2012}. In this case, the  presence of the
 solute decreases the TMD. In contrast, another two-length scale
 potential dimer model
proposed in Ref.~\cite{furlan2017} displayed a behavior in
accordance with the experimental data, but for artificially low
densities. On the other hand, atomistic simulations using either united atom
models OPLS for methanol \cite{lomba2016}, or the very recent
simulation work for
alcohol/water solutions using flexible all-atom models for methanol,
ethanol, propanol, and tert-butanol together with TIP4P/2005f
water\cite{GarciaPerez2020}, all  
 fail to reproduce the enhancement of the density
anomaly for small alcohol concentrations. In all instances the
presence of alcohol molecules induces a substantial decrease of the
TMD (up to five times larger than the experimental one for
concentrations $x_{R-OH}\sim 0.01$). 

\begin{figure}[t]
  \centering
    \subfigure[\label{dens_and_vapa}]{\includegraphics[scale=0.45]{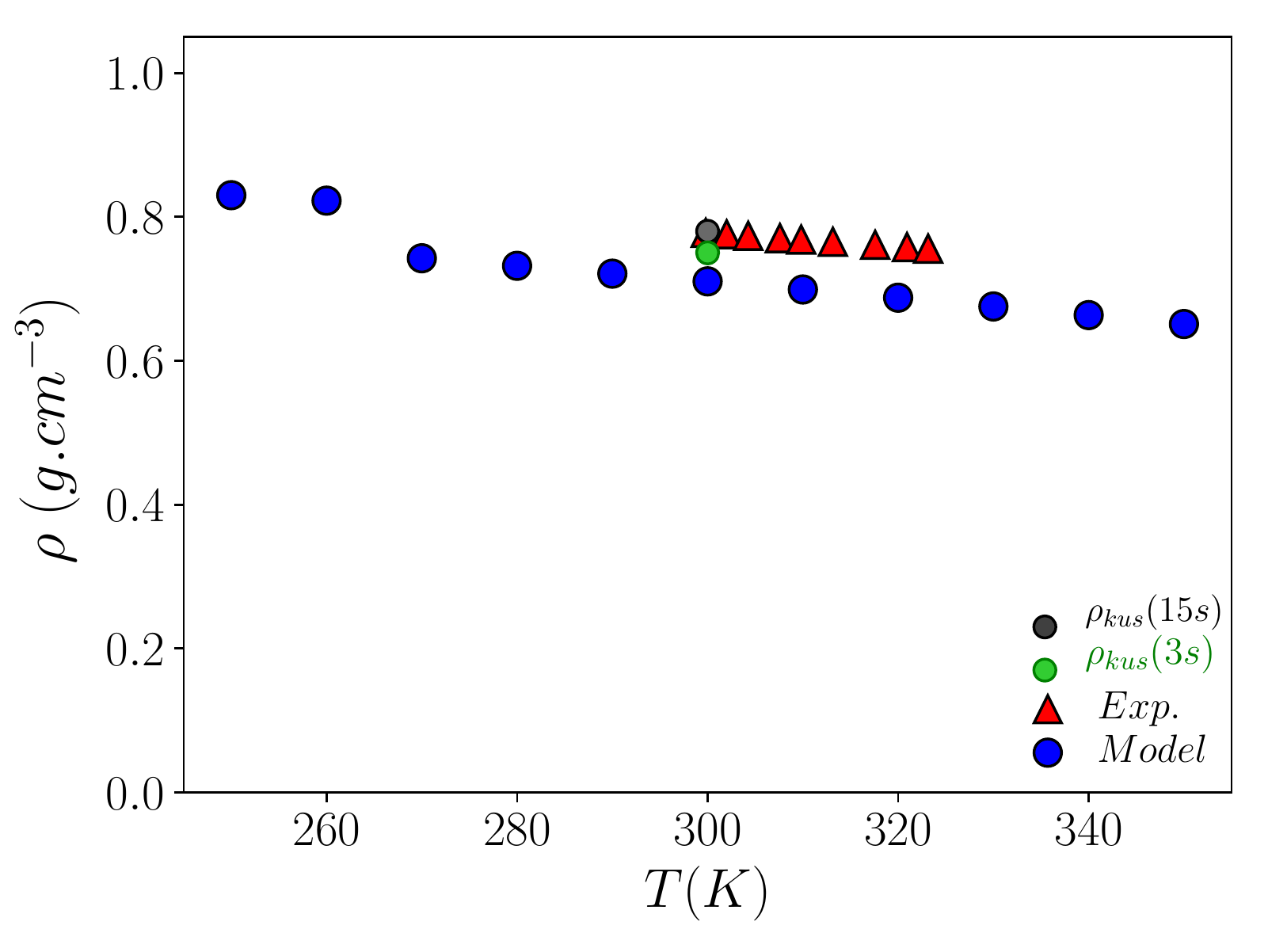}} 
    \subfigure[\label{dens_and_vapb}]{\includegraphics[scale=0.45]{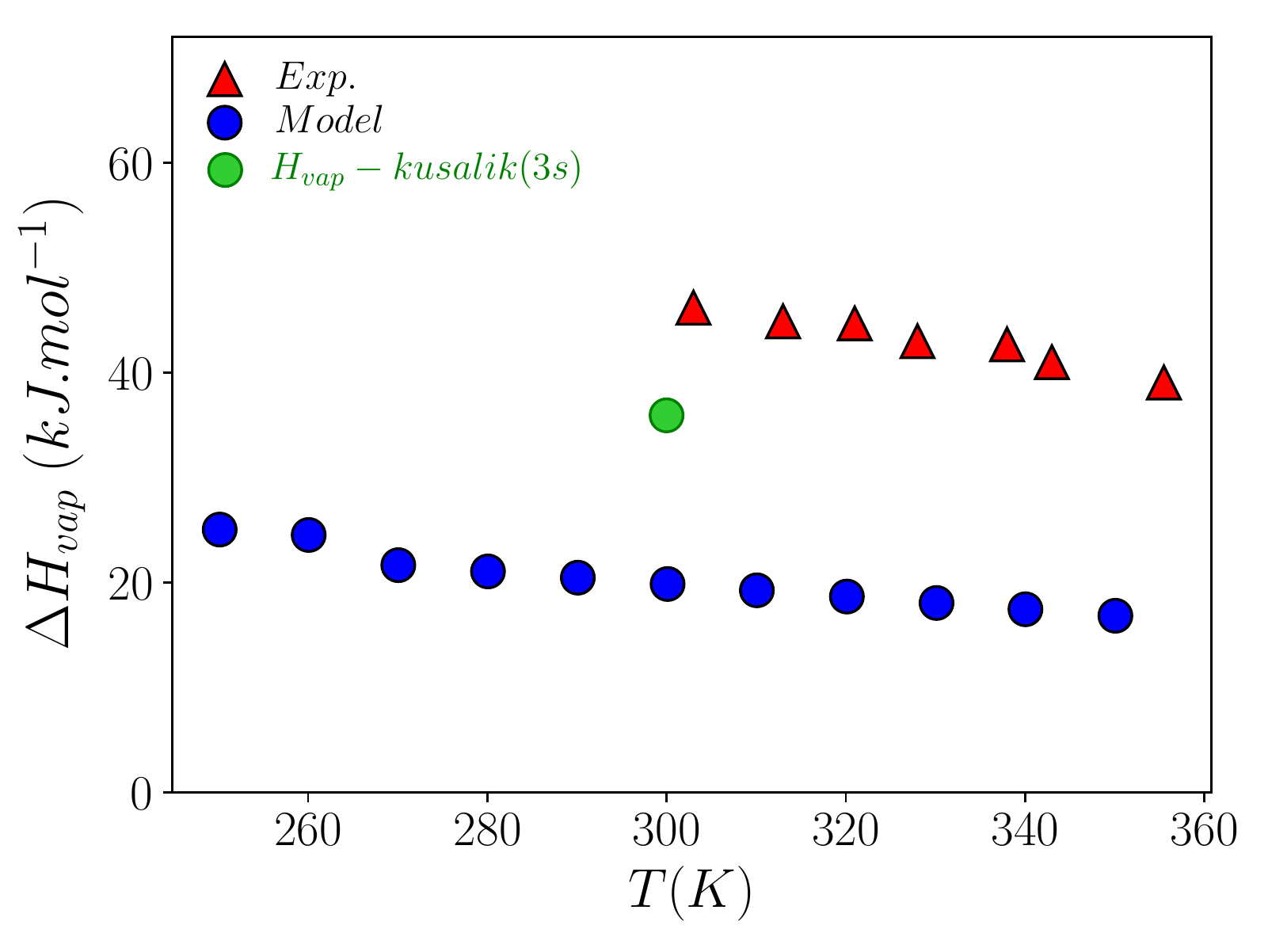}}
    \caption{(a)Density and (b)vaporization enthalpy of our model in comparison with experimental data \cite{TBAdata} and Kusalik models. 3s refers to the 3-site united atom model, and 15s to the fully atomic 15-site one \cite{kusalik2000a,kusalik2000b}.}
    \label{dens_and_vap}
\end{figure}

In this work, we will focus on tert-butanol solutions, which are of 
considerable interest due to the presence of
marked thermodynamic anomalies \cite{subramanian2011, aman2017}. For instance, highly diluted TBA solutions exhibit the most 
significant increase of the TMD among all short chain
alcohols. Moreover,  it is the highest molecular mass
alcohol to be completely miscible with water 
in all proportions under ambient conditions \cite{kasraian1995}. A 
possible origin of the anomalies has been attributed to the formation of
clathrate-hydrates \cite{kay2012}, which fits well within the
``iceberg model'' picture. Our results  will provide further evidence
in this direction. 

From a computational standpoint modeling TBA/water solutions poses a
considerable challenge. The bulky alkyl group of TBA is
known to be responsible for the formation of molecular emulsions
\cite{kezic2012}, which for certain concentrations are actually
equivalent to a microphase separation, in which two intertwined
regions of alcohol and water are separated at the microscopic
level. Accounting for these structural effects implies lengthy
simulations and extremely large samples. In addition, from the results
of Ref.~\cite{GarciaPerez2020} we know that even optimized all-atom
models do not seem capable of reproducing the experimental density
anomalies. Therefore, here we have chosen a simpler model that can
account for the structural order due to the presence of highly
directional hydrogen bonds, namely, the Stillinger-Weber 
potential, which is characterized by the presence of a strongly
directional thee-body component that favors tetrahedral coordination
\cite{stilinger1985}. This was first used to model water by Molinero and
Moore \cite{molinero2009} without explicitly accounting for hydrogen
atoms. Hereafter, we will denote
this interaction as mW potential.  We represent TBA molecules using a
two-site model, in which the alkyl group is  a
Lennard-Jones center, and the hydroxil group site interacts with other
hydroxil groups via  a modified mW
potential. Hydroxil/tert-butyl interactions are plain LJ
potentials. The tert-butyl site and hydroxil sites are 1.836 
\AA\ apart, in agreement with the geometry  parameters of Kusalik et al. three-site
model \cite{kusalik2000a}. Water will be modeled using strictly the original mW
potential, and the cross interaction between water and TBA's hydroxil
group also accounted for by a modified mW potential. A Lennard-Jones
potential is 
used to model the interaction between the alkyl group and  water. The
presence of this interaction appropriately tuned leads to an approximately correct number of hydrogen
bonds around the alcohol's hydroxil group. The parameters for the
TBA-TBA interaction are tuned to reproduce  qualitatively the
experimental density and vaporization enthalpies, at least to a comparable
level as those of the atomistic models of Kusalik et al.~\cite{kusalik2000a,kusalik2000b}.

Using extensive Molecular Dynamics (MD) simulations, we will show that
our simple model captures the increase of the TMD of 
water upon addition of small amounts of TBA. Besides, we have performed a
local structure analysis for a series of temperatures above, at and
below de TMD using Nguyen and Molinero's CHILL+ algorithm
\cite{Nguyen2014}. This algorithm allows for an identification and
quantification of ice like, clathrate, and liquid like structures in a
series of configurations of water molecules. This analysis was run on
configurations of our dimer TBA model solution and of the all atom flexible
model of Ref.~\cite{GarciaPerez2020}. In this way, we have been able
to  provides a clear
correlation between the structural reorganization at the microscopic
level and changes in the TMD. Also, one can get some insight as to 
 to why all-atom models to date do not seem capable of
reproducing the subtle effects that the presence of alcohols have on
the density anomaly.

The rest of the paper is organized as follows. In Section II we
introduce our models for water and TBA molecules, and summarize the
simulation details. Next, in Section III we discuss our most
significant results. The article is closed with a presentation of
relevant conclusions and future prospects.

\begin{figure}[ht]
\centering
\includegraphics[scale=0.42]{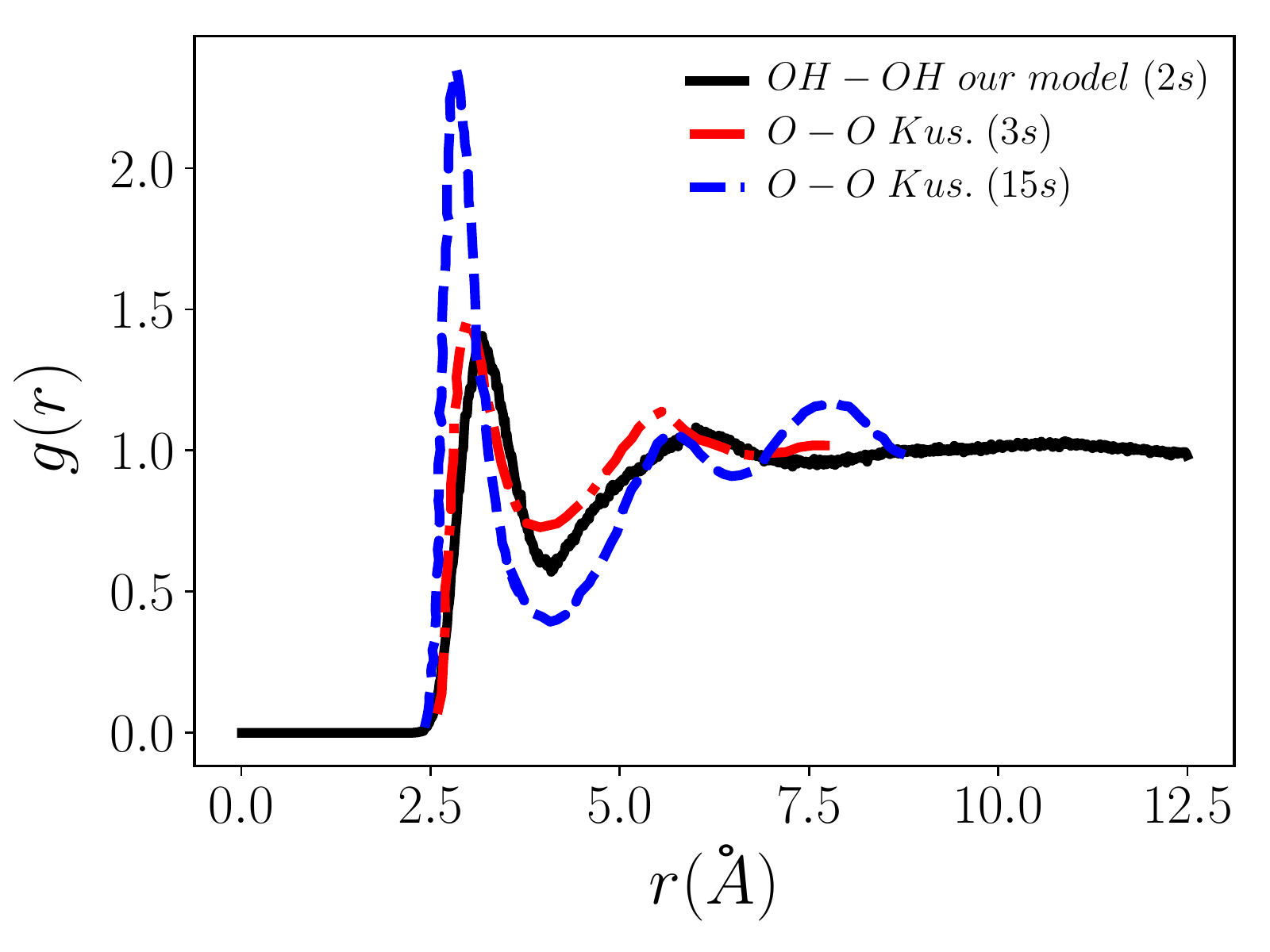}
\includegraphics[scale=0.42]{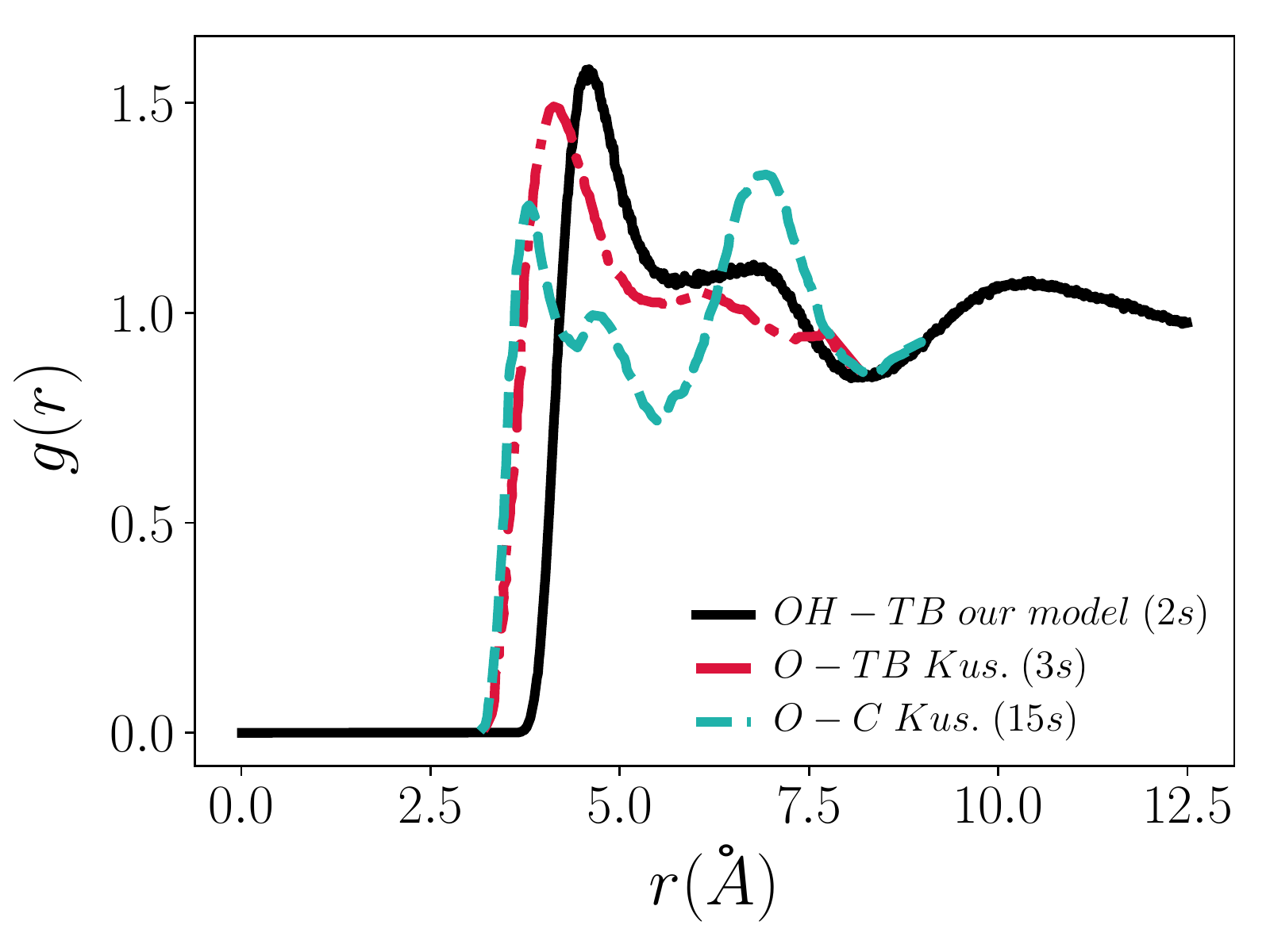}
\includegraphics[scale=0.42]{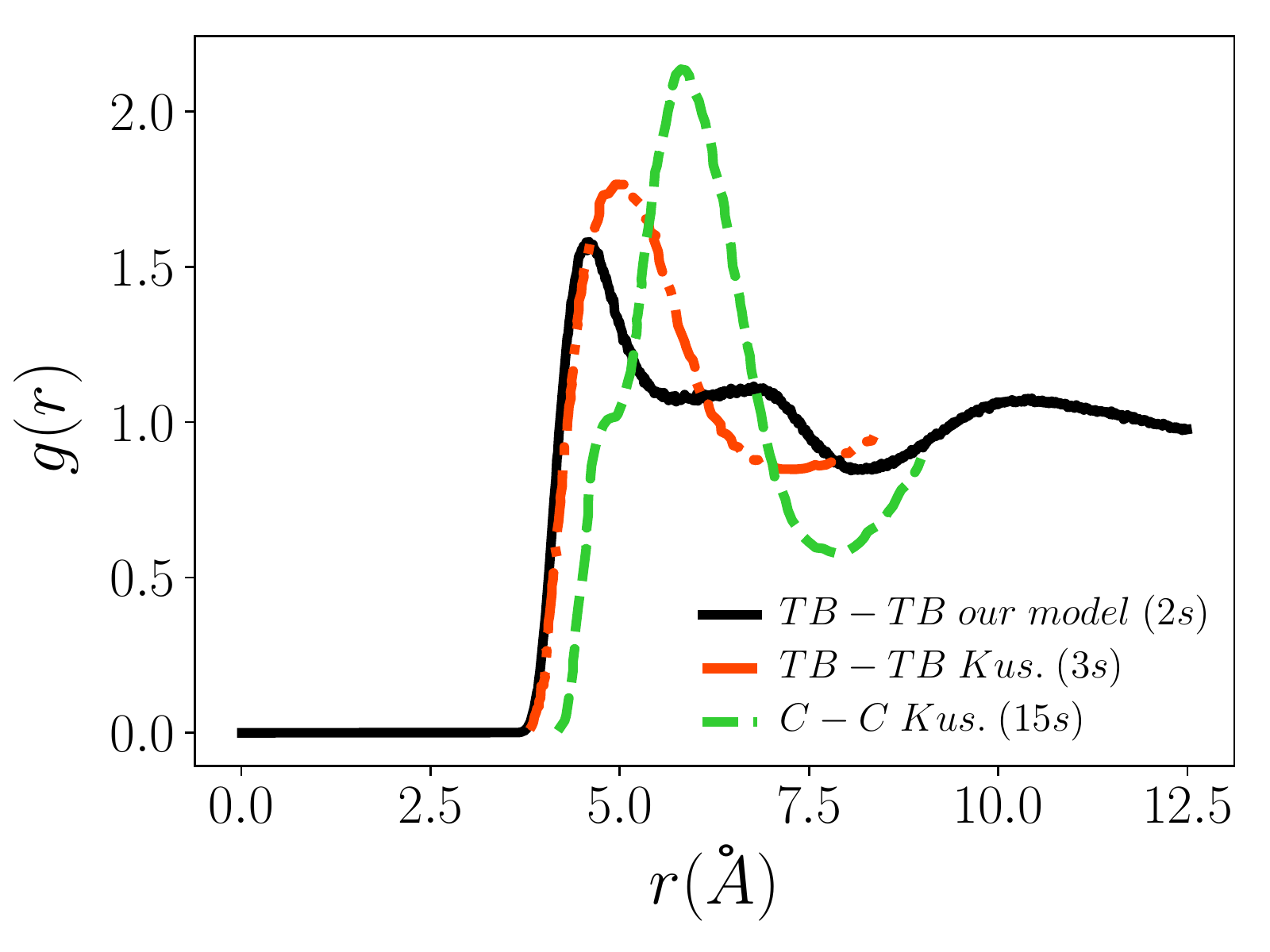}
\caption{Comparison between radial distribution function from our two-site model and Kusalik et al. two atomistic models for tert-butanol: 3-site \cite{kusalik2000a} and 15-site \cite{kusalik2000b}.}
\label{comparative_rdfs}
\end{figure}

\section{The model}
In our model we coarse-grain the 3 atom water molecule to  a single
site, and the 15 atom TBA to a two site model, as depicted in the left
graph of Figure \ref{inter} in which all the interactions involved are
represented by straight lines.

\noindent
{\bf Water.} As mentioned, for water we used the model introduced by
 Molinero and Moore \cite{molinero2009}, which was devised to tune
Stillinger-Weber's potential --originally designed for Silicon
\cite{stilinger1985} -- which guarantees the tetrahedral coordination
of oxygen atoms in ice \cite{molinero2006}. The model is a coarse-grained representation of water molecules in which only oxygen-oxygen
correlations are accounted for. It has two- and three-body
contributions of the form
\begin{equation}
    \phi(r)\;=\; \sum_{i}\sum_{j>i}\phi_2(r_{ij})+\sum_{i}\sum_{j\neq i}\sum_{k>j}\phi_3(r_{ij}, r_{ik},\theta_{ijk}),
\end{equation}
where
\begin{equation}
     \phi_2(r)=A\epsilon\left [ B\left ( \frac{\sigma}{r} \right )^p-\left ( \frac{\sigma}{r} \right )^q  \right ]exp\left ( \frac{\sigma}{r- a\sigma} \right ),
\end{equation}
and
\begin{equation}
    \phi_3(r,s,\theta)=\lambda \epsilon \left [ cos(\theta) - cos (\theta_0)\right ]^2exp\left ( \frac{\gamma \sigma}{r- a\sigma} \right )exp\left ( \frac{\gamma \sigma}{s- a\sigma} \right).
\end{equation}

The fitted parameters are $A=7.049556277$, $B=0.6022245584$, $p=4$,
$q=0$, and $\gamma=1.2$, which define the potential's form and
scale. The reduced cutoff $a=1.8$ ensures that all terms in the
potential and forces vanish at $r=a \sigma$.
Tetrahedral angles are favored by the cosine quadratic term with the
equilibrium value set to 
$\theta_0=109.47^{\circ}$. The parameter $\lambda$ tunes the repulsive
three-body term, setting  the strength of the directional 
interactions of the model \cite{molinero2009,stilinger1985}.

\noindent
{\bf Tert-butyl alcohol.} The tert-butanol molecule (See Figure \ref{inter}) is coarse-grained into
a two site model: an apolar tert-butyl center (TB) which interacts via a Lennard-Jones potential ($U_{11}$) and the hydroxyl group OH which interacts through a Lennard-Jones potential with TB centers ($U_{12}$) and  with other OH sites via a Stillinger-Weber potential similar to  the mW interaction ($U_{22}$). The parameters of the interactions were adjusted having in mind the experimental values of various thermodynamic and structural quantities: density --Fig. \ref{dens_and_vapa}, vaporization enthalpy - Fig. \ref{dens_and_vapb} and the radial distribution
functions obtained for  the three-site and fifteen-site models of Kusalik and coworkers \cite{kusalik2000a,kusalik2000b}(see Fig. \ref{comparative_rdfs}). Cross interaction parameters are determined from the standard Lorentz-Berthelot combination rules. LJ potentials are truncated at 12.5 \AA\ and long range corrections to the energy and pressure are applied.  Explicit parameter values are collected in Table \ref{TBA_parameters}. We can appreciate from Figs. \ref{dens_and_vap}
that our model performs reasonably well for the density (less than ten percent deviation from the experimental values vs. four percent of the more sophisticated three site model of Ref. \cite{kusalik2000a}) and very specially the temperature dependence is correctly reproduced. Deviations in the vaporization enthalpy are substantially larger, but in our opinion and
given the considerable departures exhibited by the more elaborate three site model, they can be deemed acceptable. Note that even for sophisticated water models, vaporization enthalpy is only reproduced when a term to account for the different self-polarization of liquid water are introduced ad hoc\cite{Pi2009}. When comparing pair distribution functions, we see that the agreement is more or less reasonable, being our model logically closer to the three site model. This is specially so for the OH-OH and OH-TB partial distributions. Differences between the fifteen site, three site and our model are in any case significant. Not surprisingly, the 15-site model yields pair distribution functions that seem to be in better qualitative agreement with experimental results from neutron diffraction \cite{Bowron1998a}.  As to hydrogen bonding, our model gives a coordination number from the integration of $g_{OH-OH}$ up to its first minimum of 2.07, which is somewhat larger than the values 1.62 and 1.77 of Kusalik et al. for three site and 15 site models respectively\cite{kusalik2000a}. Experimental estimates  lie in the range from 1.4 to 1.8 \cite{Bowron1998a}. Nonetheless, it is to be noticed that our estimate for the hydrogen bonding capabilities of our model are also well below the maximum possible value of 3. This is a clear indication of the weakening of the hydrogen bonding due to the steric hindrance induced by the bulky alkyl group.

\begin{table}[b]
\begin{tabular}{|c|c|c|c|c|c|c|c|c|}\hline
    \multicolumn{5}{|c|}{\bf OH - OH interactions} & \multicolumn{2}{|c|}{\bf TB-TB interactions} & \multicolumn{2}{|c|}{\bf TB-OH interactions}\\     \hline
    $\epsilon(kcal/mol)$ & $\sigma(\mbox{\AA})$  & $\lambda$  &
    $\gamma$ & p& $\epsilon(kcal/mol)$ & $\sigma(\mbox{\AA})$ & $\epsilon(kcal/mol)$ & $\sigma(\mbox{\AA})$ \\  \hline
    1.50 & 2.60 & 65.00 & 1.2 & 5 & 0.25 & 5.45 &
    LB  & LB \\  \hline
    \end{tabular}
    \caption{Parameters for OH-OH, TB-TB and TB-OH interactions. Cross
      interaction parameters are computed using the standard
      Lorentz-Berthelot (LB) combining rules. In
      the case of OH-OH, all remaining parameters take the original
      values of Moore and Molinero \cite{molinero2009}.}
\label{TBA_parameters}
\end{table}

\begin{figure}[t]
  \centering
  \subfigure[\label{vex}]{\includegraphics[scale=0.45]{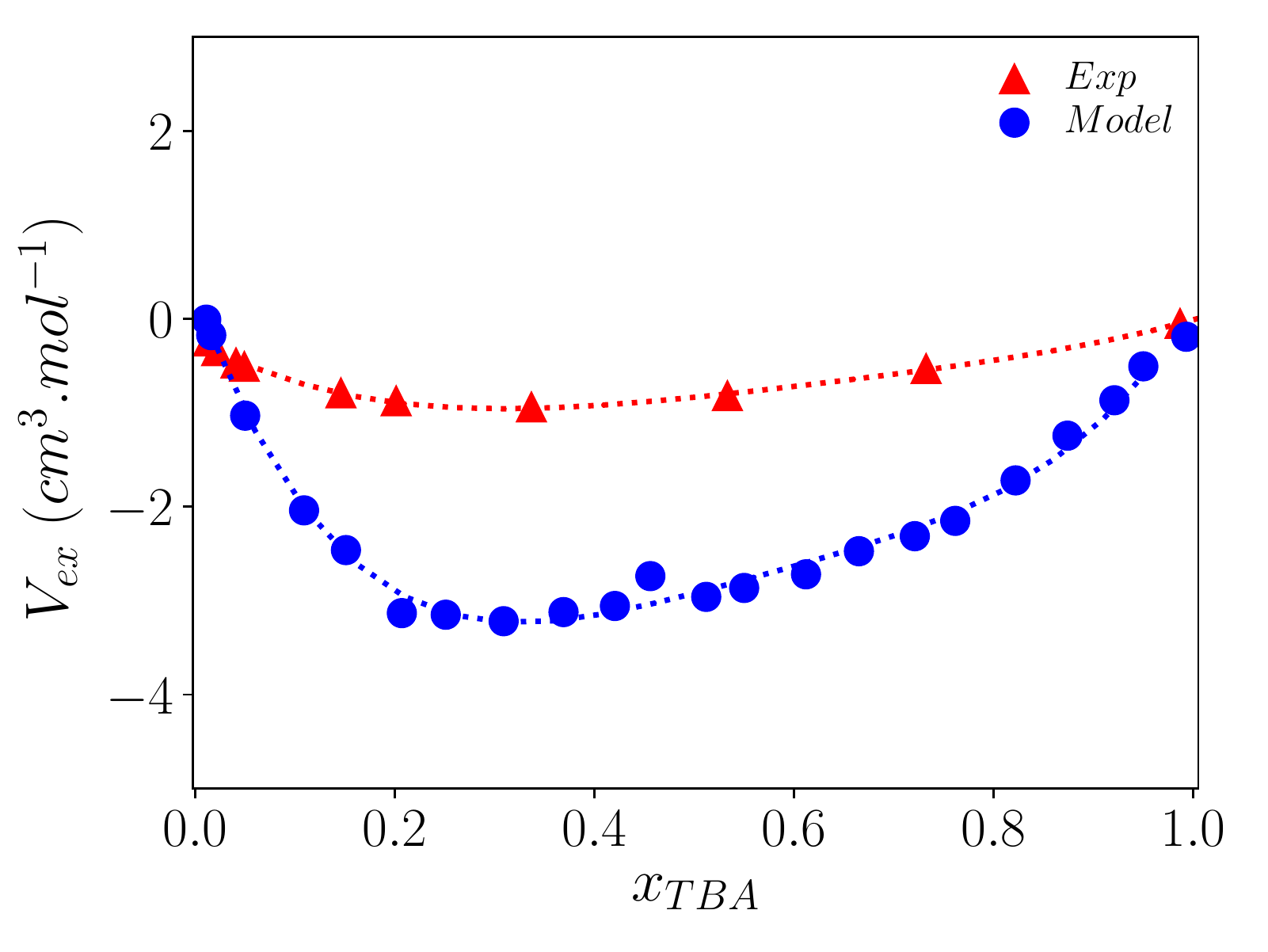}}
  \subfigure[\label{hex}]{\includegraphics[scale=0.45]{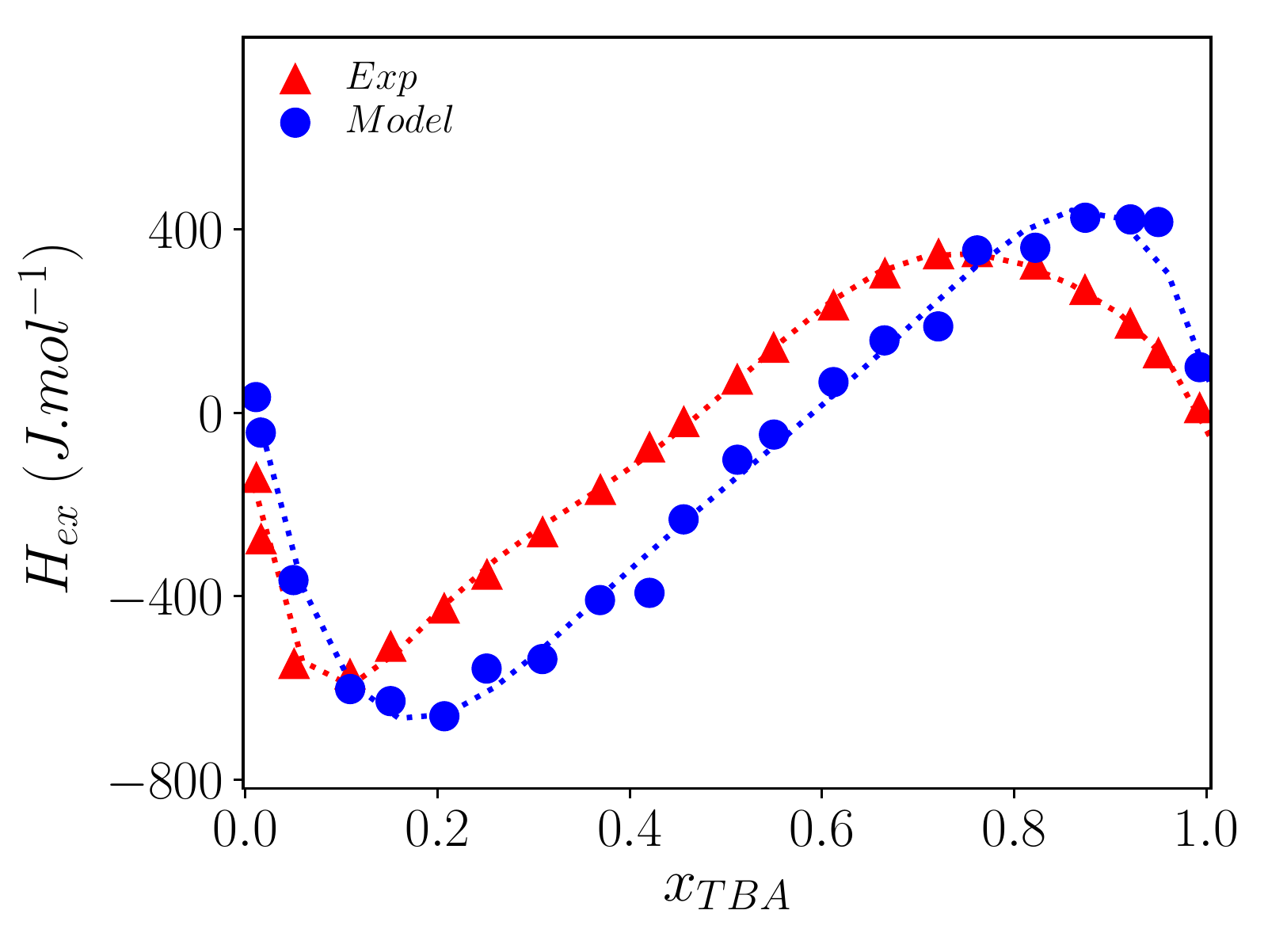}}
\caption{Experimental (red) and our model's (blue) excess thermodynamic
  properties of TBA/water solutions: excess volume (left) and excess
  enthalpy (right).\label{excess}}
\end{figure}

\noindent
{\bf Cross interactions for the TBA-water}. The effects of mixing are best analyzed in terms of excess quantities, which are nothing but the difference between mixture thermodynamic properties both real and ideal at a given composition, pressure and temperature. As shown in Ref.~\cite{lomba2016}, excess properties from simulated models can hardly be reproduced if the cross interactions between different molecular components are computed using standard mixing rules. The obvious route to bypass this shortcoming is to adjust these cross interaction parameters to fit the experimental value of the excessproperties over the whole composition range. In our case we have used
as reference quantities to be fitted the excess  enthalpy
\cite{koga1988} and excess volume \cite{Egorov2011}). Results from the
fit are illustrated in Figure \ref{excess}, and we can see that
the model reproduces qualitatively the experimental behavior, both
the volume contraction and the non-monotonic compositional dependence
of  the excess enthalpy. Tert-butyl/water interaction ($U_{13}$) is modeled via a plain LJ potential truncated at 12.5 \AA\ with long range corrections to the energy and pressure are applied. The OH-water interaction ($U_{23}$) is again a Stillinger-Weber potential with a three-body component, for which the $\epsilon$ and $\sigma$ parameters have been optimized and the remaining parameter are identical to those of the plain mW water-water interaction. The final fitted parameters are collected  in Table \ref{table_cross_parameters}.

\begin{table}[b]
    \centering
\begin{tabular}{|c|c|c|c|} \hline
\multicolumn{2}{|c|}{\bf TB-mW interactions} & \multicolumn{2}{|c|}{\bf OH-mW interactions}\\     \hline
   $\epsilon$ (kcal/mol)  &  $\sigma(\mbox{\AA})$ & $\epsilon$(kcal/mol) & $\sigma(\mbox{\AA})$ \\ \hline
    0.459 & 3.984 & 1.371 & 3.660\\ \hline
\end{tabular}
\caption{Optimal cross parameters for our TBA-water mixture model. In
      the case of OH-mW, all remaining parameters take the original
      values of Moore and Molinero \cite{molinero2009}.}
    \label{table_cross_parameters}
\end{table}

\begin{figure}[ht]
  \centering
  \includegraphics[scale=0.7]{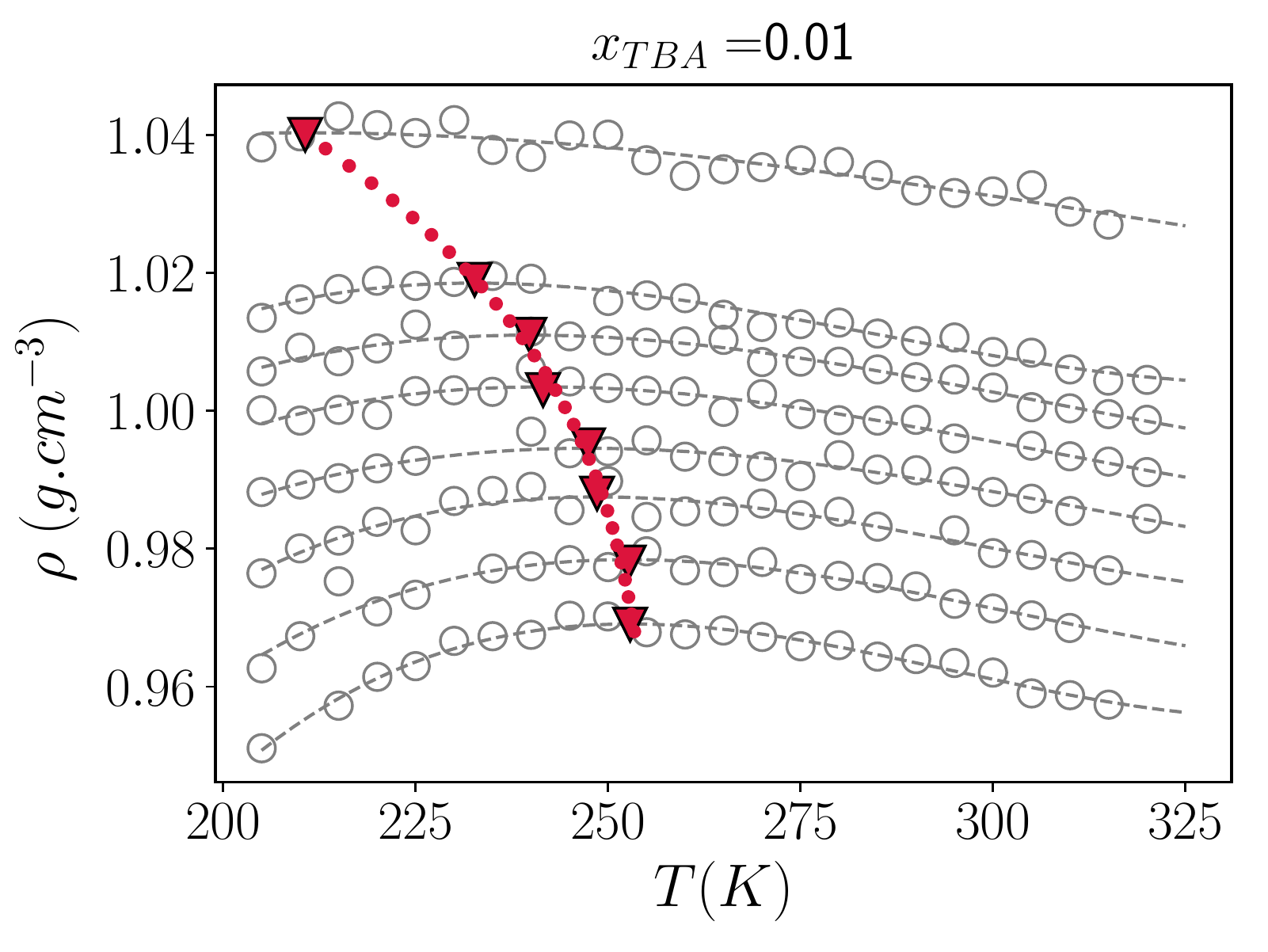}
  \caption{Density isobars for pressures, 1 bar, 500 bar,
    1000bar,...5000bar (from bottom to top) for TBA in water with
    $x_{TBA}=0.01$. Simulation data are denoted by 
    symbols and lines correspond to a third degree polynomial fit.\label{rhoX}}
\end{figure}

\begin{figure}[t]
  \centering
  \includegraphics[scale=0.7]{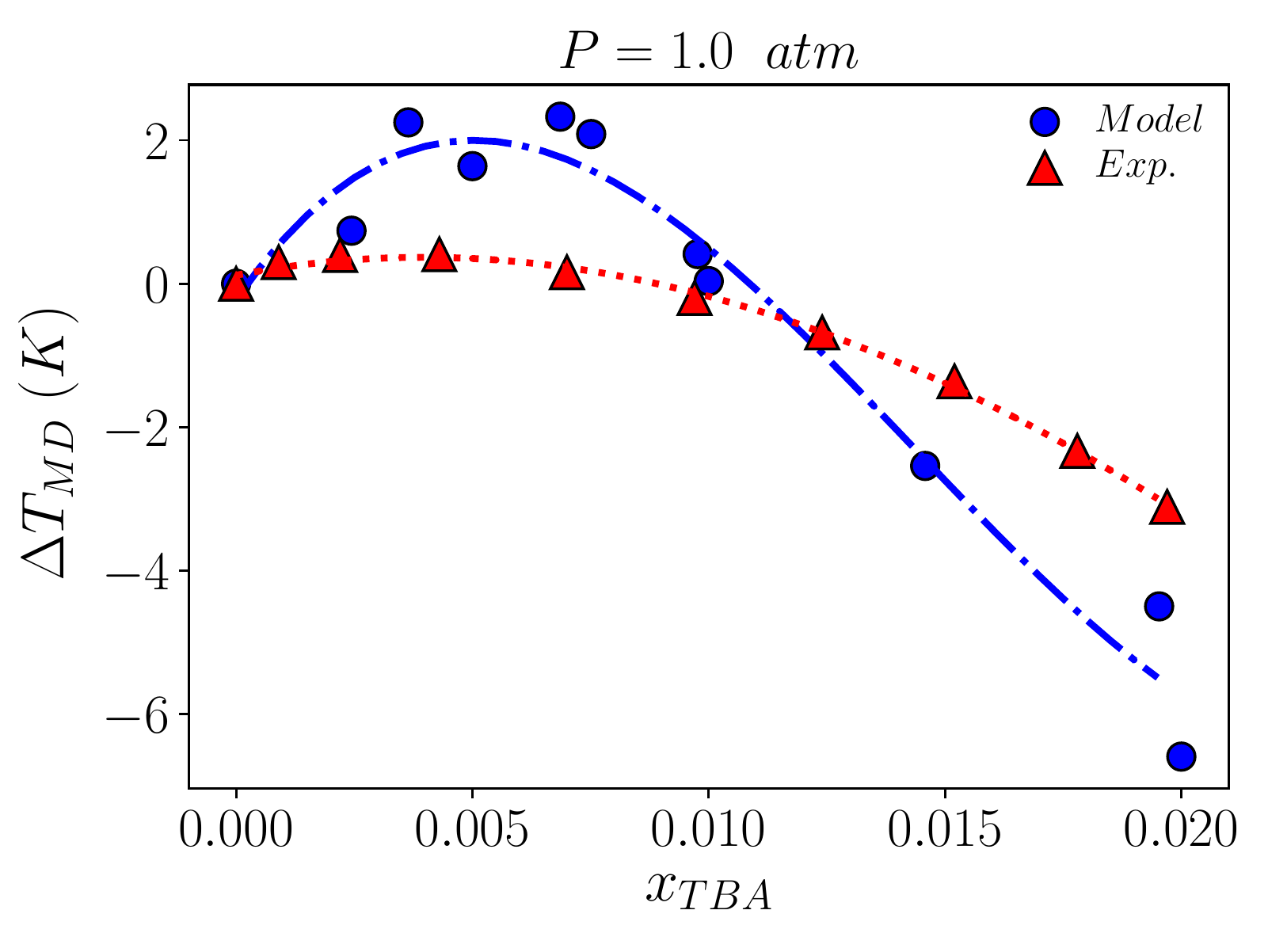}
  \caption{Mole fraction dependence of the change in the TMD of
    water/TBA solutions with respect to that of pure water. Red
    triangles and dash-dotted curve correspond to our model results, blue
    triangles and dashed curve  denote experimental data.\label{delTMD}}
\end{figure}
\begin{figure}[t]
  \subfigure[\label{histoMw}]{\includegraphics[clip,scale=0.40]{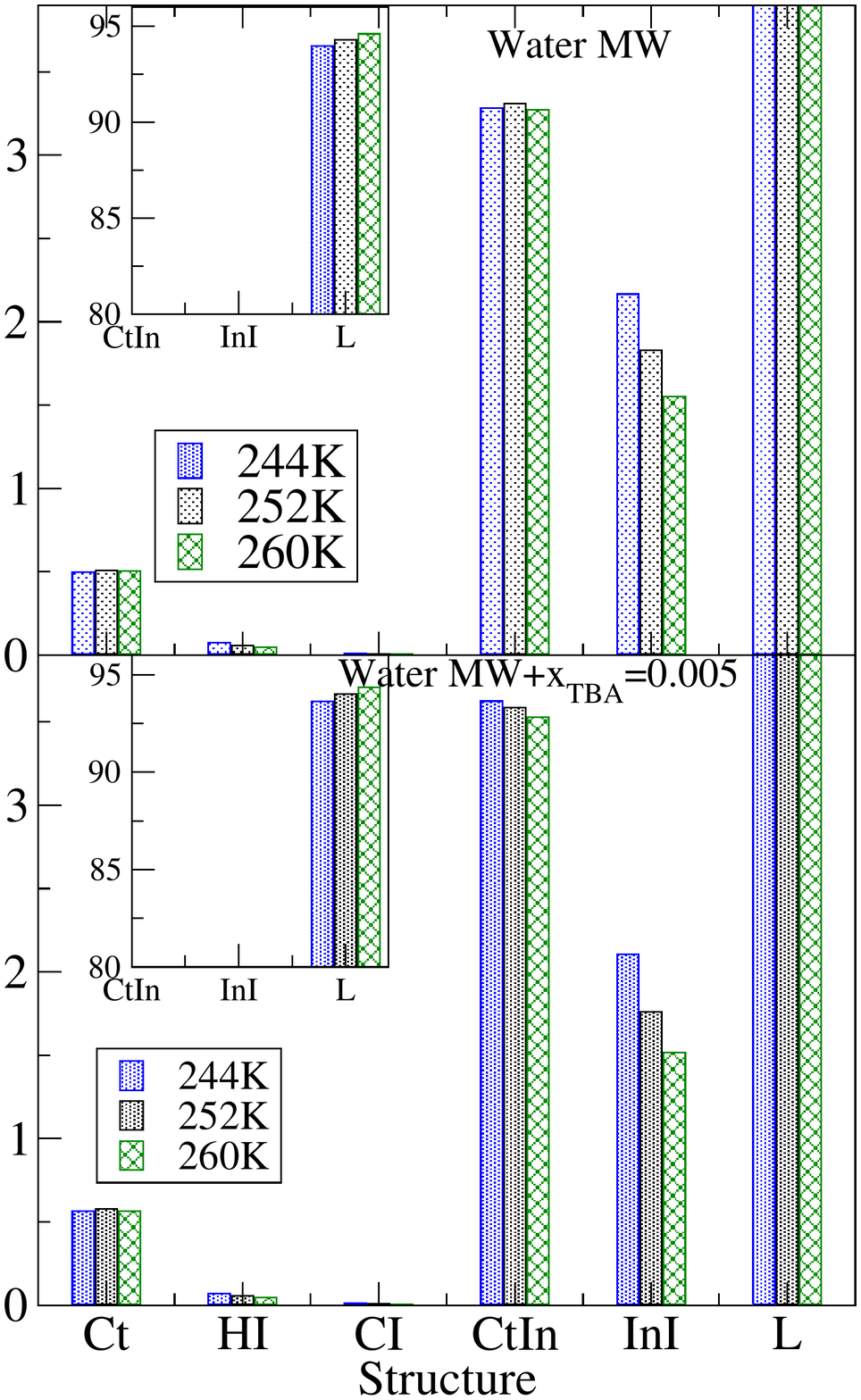}}
  \subfigure[\label{histoTip}]{\includegraphics[clip,scale=0.40]{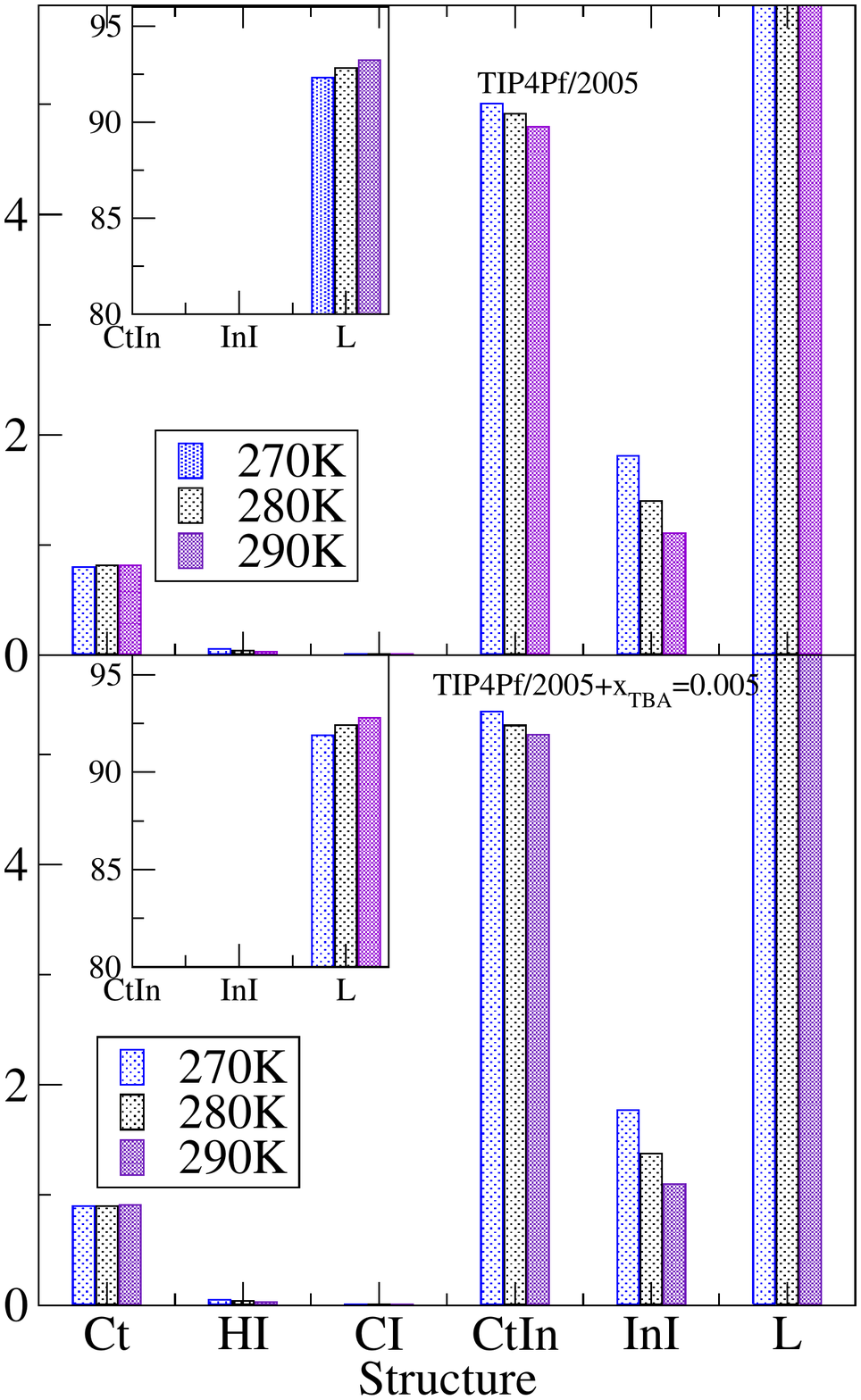}}
  \caption{Histograms of relative abundance of local
    clathrate hydrates (Ct), hexagonal ice (HI), cubic ice (CI),
    interfacial clathrates (CtIn), interfacial ice (InI), and liquid
    water (L), as determined using CHILL+ algorithm\cite{Nguyen2014}
    on 2000 configurations of our TBA/water model (lower left graph), and
    the flexible all-atom TIP4P/2005f-OPLS model of
    \cite{GarciaPerez2020} (lower right graph). The corresponding
    systems for pure mW and TIP4P/2005f water are depicted in the
    upper graphs\label{histoT}}
\end{figure}

We have performed MD simulations for a number of system with  
particle numbers  ranging from 2000 (pure water) to 4000 (pure TBA) for
various compositions using the LAMMPS package \cite{plimpton1995}. 
Simulations were performed in the isothermal-isobaric ensemble with a Nosé-Hoover
thermostat and barostat \cite{nose1984,hover1985} with a time-step of
one fs and relaxation times of 10ps and 100ps respectively. Particles were
placed in a cubic box with standard periodic boundary condition. The
dimer bonds were kept fixed using a SHAKE algorithm
\cite{ryckaert1977}, with a tolerance factor of $10^{-4}$. Our
simulations started from a compositionally disordered mixture of TBA
and water particles, which was equilibrated at the chosen pressure and
temperature for 2 ns. Production runs were 10 ns
long. To ensure that the system was thermalized, the evolution of the pressure,
and the kinetic and potential energies were closely monitored during the
equilibration run. Configurations  were stored every 2 ps and running
averages computed every 0.1 ps. Additionally, we have run all-atom
simulations using an optimized OPLS-AA model proposed by Jorgensen et
al. \cite{Jorgensen1996} in combination with a TIP4P/2005f flexible
model for water \cite{Gonzalez2011} and cross interaction parameters
fitted to experimental excess properties
\cite{GarciaPerez2020}. Additionally simulations for pure TIP4P/2005f
water were also run.  We have used the GROMACS package \cite{gromacs,Abraham2015}
in the isothermal-isobaric ensemble with a time-step of 0.5
fs. Configurations were stored every 2000 time-steps for temperatures
approximately at the TMD  and some 10K above and
below, in order to analyze the structural changes taking place when
crossing the temperature of maximum density at constant pressure. 

\section{Results and Conclusions}
After a long series of simulations for TBA mole fractions $x_{TBA}\le
0.05$ at various pressures, we have obtained density isobars as those
shown in Figure \ref{rhoX}, in which the dashed curves correspond to
third degree polynomial fits. We have used these fits to obtain the
estimates of the temperature of maximum density. Despite some
statistical uncertainty in the simulated densities (due to the high
accuracy required to appreciate the very small density change), the trends are
clear. One observes that as pressure grows  the TMD decreases in a
monotonic fashion.  This is in agreement with experimental findings \cite{Pi2009},
and it occurs for all the mole fractions studied for which the anomaly
is preserved. The origin of this behavior can be traced back to the
fact that pressure tends to hinder the formation of low density
ice-like and clathrate structures by which the anomalous region of water (or water
solutions here) is shifted to lower temperatures. 

Now, the relevant quantity is precisely the change in the TMD due to
the addition of solute, $\Delta T_{MD}(x_{TBA})$. This quantity is
plotted in Figure \ref{delTMD} as computed from the simulation
results, together with the experimental results de Wada and Umeda
\cite{wada1962a}. The first feature one observes in the figure is the
presence of a 2K maximum at $\sim x_{TBA}=0.005$, which is
approximately the location of the maximum for the experimental
results. Our model overestimates the temperature change of the maximum. After
the maximum, one rapidly  reaches $\Delta T_{MD}(x_{TBA})<0$ for $x_{TBA}\approx 0.01$, a
value slightly higher than that of the experimental crossover. For
larger concentrations the TMD decreases further, as it does experimentally, up
to a point where it is either destroyed (TBA does not exhibit any density
anomaly) or preempted by crystallization.

Thus, we have finally devised a simple model of short chain alcohol that
displays the ``structure maker'' character observed experimentally,
i.e. a solute that increases the temperature of the maximum density of
water. Interestingly,  in a very recent work \cite{GarciaPerez2020} a
much more sophisticated model for water/TBA mixtures using flexible
all-atom simulations was unable to reproduced this feature. In order
to correlate the non-monotonic density dependence of water and
water/TBA with  microscopic structural changes, we have analyzed a
series of configurations from our TBA/water model and the
TIP4P/2005f-OPLS of Ref.~\cite{GarciaPerez2020} using Nguyen and
Molinero's CHILL+ algorithm \cite{Nguyen2014}. This procedure allows for the
identification and quantification of cubic ice, hexagonal ice, and
clathrate hydrate structures (both bulk and interfacial) together with
liquid water ones. Results for both models approximately at the TMD
and 8-10K above and below are shown in Figs~\ref{histoT}, in addition to
pure water results modeled using mW potential \cite{molinero2009} and
the flexible TIP4P/2005f model \cite{Gonzalez2011}. Statistical
uncertainties are not visible at the scale of the figure. Focusing first on
water, one observes that only clathrate hydrate structures exhibit a subtle
maximum at the TMD for both water models.  In the case of the mW
potential a maximum is also present in interfacial clathrate structures. All
these structures are tetra-coordinated oxygen atoms but with eclipsed bonds
\cite{Nguyen2014}. As temperature increases, both these and
liquid-water structures grow initially at the expense of ice-like
structures. Being almost perfectly tetrahedral, ice-like structures are
less dense. This explains the initial anomalous increase of density. From
the TMD onward, the relative weight of low density structures
(clathrate, and ice-like structures both bulk and interface)
diminishes considerably, and density decreases due to the regular
thermal expansion of 
the high density liquid water structure.  Thus, the interplay between a small
maximum in low density structures (clathrates) and increase of
liquid-like (high
density) structures seems to be at the source of the existence of
a TMD. Note, additionally, that all other low density (ice-like)
structures have a very small presence after melting, and 
display a monotonic decrease with increasing temperature. It is
important to notice that the clathrate structures occur in both models
of pure water, so its existence does not require --although, as shown below, it is
enhanced by-- the presence of
solute molecules.

Now in the
lower graphs of Figure \ref{histoT} we have the corresponding
histograms for the solutions at $x_{TBA}=0.005$, i.e. close to the
maximum of $\Delta T_{MD}(x_{TBA})$  for our model.
 We see that again our results exhibit a
maximum in the bulk clathrate hydrate structures. Interestingly the
maximum does not occur for the
interfacial clathrates anymore. In the solution, these structures are
basically  promoted by the presence of  solute molecules, and their
relative weight monotonically decreases with temperature. Now, in the 
TIP4P/2005f-OPLS model the maximum is shifted to temperatures well
beyond the TMD, the region shown in the figure displaying a slight
increase in the relative weight of the clathrate structures.   This explains why
this model preserves the density anomaly, but on the other hand the maximum in
the low density structures occurs now for higher temperatures, above
the TMD, where the thermal expansion of the dominant liquid-like
structures control the behavior, thus the water anomaly is not
enhanced and  $\Delta T_{MD}(x_{TBA})<0$. 

In our model, for a larger concentration, such as
 $x_{TBA} = 0.02$  the density anomaly occurs at
very low temperatures, where the large fluctuations in the results are
likely connected with the onset of crystallization and $\Delta
T_{MD}(x_{TBA})<0$. 
In this case, the solute behaves now as a strong ``structure breaker''. We have observed that the maximum in the ratio of
bulk clathrate structures is much less marked. Also, the relative weight
of interfacial clathrates
increases 25 percent with respect to the value for pure water and
diminishes with increasing temperature.  As in the case of the
all-atom model, these structures are promoted by the presence of
solute molecules and their relative weight depends on the
concentration of the latter. The decrease in the maximum of bulk
clathrate structures is most likely connected with the fall in the
TMD. Larger increases in $x_{TBA}$ will lower the TMD even 
further, and the density anomaly  will be completely preempted by
crystallization/vitrification. Apparently,  in the all-atom model
solution, the shift of the bulk clathrate structure maximum, and in
our dimer TBA model the smoothing of the corresponding maxima for
concentrations above  $x_{TBA}\sim 0.01$,  are the structural
features that determine the ``structure breaker'' character of the
solutes. Note that for lower concentrations, our model solute becomes
``structure maker''.

In summary, we have devised a simple diatomic model for TBA with three
body interactions on the hydroxil site that mimics the formation
hydrogen bonds. Cross interactions were fitted to  qualitatively
account for the experimental excess properties of water/TBA
solutions, with water represented by Molinero and Moore's model
\cite{molinero2009}. We have seen that the model is capable of
reproducing the experimental enhancement of the density anomaly of
water observed for very small concentrations of alcohol. A structural
analysis of the simulation results illustrates the correlation between
the presence of a maximum of certain clathrate structures and the
density anomaly. The fact that the maxima occurs in the bulk
clathrates and not interfacial clathrates and that high density liquid
like structures also increase in a monotonic fashion with temperature
seems to be at the root of the density anomaly
enhancement. As found in Ref.~\cite{GarciaPerez2020},  a much more
sophisticated all-atom model is unable to reproduce the experimental
behavior.

Future work should address the inability of all-atom models to account
for the concentration dependence of water anomalies. In this regard,
the role of the hydrogen bond network has to be reassessed and
possibly include non-additive effects such as polarization and charge
transfer.

\section*{Acknowledgments}
EL, EGN and DGS  acknowledge the support from the Agencia Estatal de Investigación and Fondo Europeo de Desarrollo Regional (FEDER) under grant No. FIS2017-89361-C3.  MSM and MCB thanks the  Brazilian science agencies - Conselho Nacional de Desenvolvimento Cientıfico e Tecnologico (CNPq) and  Coordenacão de Aperfeiçoamento de Pessoal de Nível Superior (CAPES Print Program) for the support to the collaborative period in the Instituto de Química Fisica Rocasolano. DG acknowledge to the Galicia Supercomputing Center (CESGA) for  the computer time allocated for some of our  calculations.

\section*{Data Availability}
The data that support the findings of this study are available from the corresponding author upon reasonable request.

\bibliography{myref}
\bibliographystyle{ieeetr}

\end{document}